\documentclass[a4paper]{jpconf}
\usepackage{graphicx}
\usepackage{amsmath}
\usepackage{amssymb}

\newcommand{\Id}{{\rm Id}}
\newcommand{\SU}{{\rm SU}}
\newcommand{\U}{{\rm U}}
\newcommand{\Const}{{\rm Const.}}
\newcommand{\supp}{{\rm supp}}
\newcommand{\im}{{\rm Im}}

\newcommand{\sgn}{{\rm sgn}}
\newcommand{\Ker}{{\rm Ker}}
\newcommand{\ext}{{\rm ext}}
\newcommand{\eps }{\varepsilon}
\begin{document}
\title{The one dimensional semi-classical Bogoliubov-de Gennes Hamiltonian
with PT symmetry: generalized Bohr-Sommerfeld quantization rules}

\author{A~Ifa}

\address{Universit\'e Tunis El-Manar, Tunis, Tunisia}

\author{M~Rouleux}

\address{Aix-Marseille Univ, Universit\'e de Toulon, CNRS, CPT, Marseille, France}

\ead{abdelwaheb.ifa@fsm.rnu.tn; rouleux@univ-tln.fr}

\begin{abstract}
We present a method
for computing first order asymptotics of semiclassical spectra for 1-D Bogoliubov-de Gennes (BdG) Hamiltonian 
from Supraconductivity, which models the electron/hole scattering through two SNS junctions. 
This involves: 1) reducing the system to Weber equation near the branching
point at the junctions; 2) constructing local sections of the fibre bundle of microlocal solutions; 3) normalizing these
solutions for the ``flux norm'' associated to the microlocal Wronskians; 4) finding the relative monodromy matrices
in the gauge group that leaves invariant the flux norm; 5) from this we deduce Bohr-Sommerfeld (BS) quantization rules
that hold precisely when the fibre bundle of microlocal solutions (depending on the energy parameter $E$)
has trivial holonomy.
Such a semi-classical treatement reveals interesting continuous symetries related to monodromy. Details will appear elsewhere.
\end{abstract}

\section{\bf Bogoliubov-de Gennes Hamiltonian}
\smallskip
BdG Hamiltonian describes the dynamics of a pair of quasi-particles electron/hole in the Theory of Supraconductivity \cite{BCS}. 
We consider a narrow metallic 1-D wire (Normal Metal N) connected to Supraconducting bulks S through a 
SNS junction, and compute 
the excitation spectrum in the normal contact region as a function of gate voltage, when electronic levels transform into phase sensitive
Andreev levels.
The wire, or lead, is identified with a 1-D structure, 
the interval $x\in[-L,L]$ (case of a perfect junction) or $x\in[-L+\ell/2,L-\ell/2]$ (``dirty junction''), where $\ell\ll L$.
The reference energy in the lead is Fermi level $E_F$. The pair electron/hole is acted upon by two kinds of potentials:

(1) the ``order parameter'' $\Delta(x)$ times a phase function $e^{i\phi(x)/2}$, which is the potential due to Cooper pairs in the supraconducting 
bulk. This potential, subject to self-consistency relations, is priori unknown. Namely, 
inside S, $\Delta(x) e^{i\phi(x)/2}$ is a solution of Ginzburg-Landau (or Pitaevskiy) equations, and shows typically a vortex profile (in 2-D). 
In BdG Hamiltonian it is assumed, however, that $\Delta(x) e^{i\phi(x)/2}$ is an ``effective'' potential.
Inside N, superconducting gap $\Delta(x)\equiv0$: quasi-particles live in the ``clean metal''. For $|x|\geq L+\ell$, $\Delta(x)=\Delta_0>0$. 

We assume that the phase function $\phi(x)$ is constant near the junction, and
gauge the interaction  by $\phi_-=-\phi_+=-\phi$ in the superconducting banks, so that
$\phi(x)=\mathrm{sgn}(x)\phi$. We assume further that this equality holds everywhere: since $\Delta(x)=0$ inside N, the discontinuity 
of $x\mapsto\phi(x)$ is irrelevant.

(2) a smooth chemical potential $\mu(x)$: typically $\mu(x)$ is flat in N and drops smoothly to the band bottom in the 
superconducting banks S. In our model we assume again $\mu(x)$ to be constant in the superconducting bank, i.e. $\mu(x)=\mu_0$
when $|x|\geq L+\ell$. Andreev currents at energy $E$ occur only if $\mu(x)\geq E$ in $[-L,L]$.

The case of a perfect junction ($\Delta$ ``hard-wall potential'') has been considered in \cite{ChLeBl},
see also \cite{CaMo} for a SFS junction, and makes use
scattering matrix techniques. In this work, justifying semi-classical techniques
as in \cite{DuGy} (also in the multi-dimensional case)
we rather consider an imperfect (or ``dirty'') junction: $\Delta(x)e^{i\phi(x)/2}$  
is a smooth function. 
In a neighborhood of $[-L,L]$, say $x\in[-L-\ell,L+\ell]$, the system is described 
at the classical level by BdG Hamiltonian 

\begin{equation}\label{1}
{\cal P}(x,\xi)=\begin{pmatrix}\xi^2-\mu(x)\kern 1pt &  \Delta(x)e^{i\phi(x)/2}\kern 1pt \\
\Delta(x)e^{-i\phi(x)/2}\kern 1pt & -\xi^2+\mu(x)\kern 1pt\end{pmatrix}
\end{equation}

The energy surface:
$\Sigma_E=\{\det({\cal P}-E)=-(\xi^2-\mu(x))^2-\Delta(x)^2+E^2=0\}=\Lambda_E^<\cup\Lambda_E^>$
splits into 2 branches separated in momentum space, so consists of two microlocal wells. Interaction between these 
wells gives the imaginary parts of the resonances for the electron/hole scattering, and will be ignored in this paper.
Because of smoothness of $x\mapsto\Delta(x)$, the reflections occur inside $[-L,L]$, we denote by $(\pm x_E,\xi_E)\in\Lambda_E^>$,
the one-parameter family of ``branching points''
defined by $\Delta(\pm x_E)=E$ with $x_E$ near $x_0\in[L-\frac{\ell}{2},L+\frac{\ell}{2}]$, $\Delta(x_0)>0$.
We do not consider the problem of ``clustering'' of eigenvalues as $E\to0=E_F$ (Fermi level).
In the ``hard wall potential'' limit for $x$ near $x_0$,
the potential $\Delta(x)$ can be safely approximated by a linear function such that $\Delta(x_0)=E_0$,
and $\mu(x)$ by a constant $\mu$.
So near $x_0$ we assume that 
\begin{equation*}
\phi(x)=\phi, \quad \mu(x)=\mu>E, \quad \Delta(x)=E+\alpha(x-x_E)
\end{equation*}
for large $\alpha>0$. Condition $a_E=(x_E,\xi_E)\in\Sigma_E$ gives $\xi_E^2=\mu>E$, $\Delta(x_E)=E$.

The physical mechanism goes roughly as follows (see \cite{ChLeBl} for a detailed exposition):
An electron $e^-$ moving in the metallic lead, say, to the right,  with energy $0<E\leq\Delta$ below the gap
and kinetic energy $K_+(x)=\mu(x)+\sqrt{E^2-\Delta(x)^2}$ is
reflected back as a hole $e^+$ from the supraconductor, injecting a Cooper pair into the superconducting
contact.  The hole has kinetic energy $K_-(x)=\mu(x)-\sqrt{E^2-\Delta(x)^2}$, and a momentum of the same sign as
this of the electron.
When $\inf _{[-L,L]}K_-(x)>0$ it
bounces along  the lead to the left and picks up a Cooper pair in the supraconductor, transforming again to the
original electron state, a process known as Andreev reflection.
This works also the other way in $\Lambda_E^<$,  since Hamiltonian system conserves both charge and energy. 
Actually, the hole can propagate throughout the lead  only if $\inf _{[-L,L]}\mu(x)\geq E$.
Otherwise, it is reflected from the potential $\mu(x)$ in the junction, and Andreev levels are quenched at higher energies,
i.e. transform into localized electronic states.

For a rescaled ``Planck constant'' $h$ so that $h\ll\ell$,
we consider Weyl $h$-quantization of  BdG  Hamiltonian
${\cal P}(x,hD_x)$ on $L^2(I)\otimes{\bf C}^2$, $I=[-(L+\ell),L+\ell]$,
which is self-adjoint when imposing Dirichlet boundary conditions at $\partial I$. 
Phase-sensitive Andreev states carry supercurrents that turn out to be proportional
to the $\phi$-derivative of the eigen-energies of ${\cal P}(x,hD_x)$.

We have
$\sigma^y{\cal P}(\phi)\sigma^y=-{\cal P}(-\phi)$, with $\sigma^y=\begin{pmatrix}0&-i\\i&0\end{pmatrix}$,
accounting for ``negative energies''. We shall assume here $E>0$.
When potentials are even functions (typical for metals), ${\cal P}(x,hD_x)$ verifies PT symmetry
${}^\vee{\cal I}{\cal P}(x,hD_x)={\cal P}(x,hD_x){\cal I}{}^\vee$
which is essential for our approach to work. 

At least formally, since BdG is only defined locally near N,
removing boundary conditions leads to ``resonances'' (i.e. metastable states or quasi-particles with a finite life-time).
Thus for simplicity we have assumed that (\ref{1}), together with its semi-classical quantization,
describes the system not only in $I$, but on the whole real line, provided $h\ll\ell\ll L$. Thus 
${\cal P}(x,hD_x)$ extends to $L^2({\bf R})\otimes{\bf C}^2$,
 
Our general goal is to  give a precise mathematical meaning to these ``resonances''. 
Here we content to compute their real parts through Bohr-Sommerfeld quantization rules.

\section{Monodromy operator, scattering matrix: an outlook}
\medskip
\noindent {\it a) Schr\"odinger operator on the real line.}

\smallskip
We first recall from \cite{Ar2} basic facts for a 1-D Schr\"odinger operator with a compactly supported potential $V$. 
The generalized wave-functions $u$ with energy $E=k^2>0$ satisfy
\begin{equation}\label{2}
-h^2 u''(x)+V(x)u(x)=Eu(x)
\end{equation}
and outside supp $V$,
\begin{equation}\label{3}
-h^2 u''(x)=k^2u(x)
\end{equation}
defines the state space ${\cal Z}\approx{\bf C}^2$ of the ``free particle'', spanned by $f_1(x)=e^{ikx/h}$, $f_2(x)=e^{-ikx/h}$. 
The monodromy operator $M(k):f_1+Bf_2\mapsto Af_1$ is such that
\begin{equation*}
M(k)=\begin{pmatrix}1/\overline A&-\overline B/\overline A\\-B/A&1/A\end{pmatrix}\in \SU(1,1)
\end{equation*}
In particular, $|A|^2+|B|^2=1$. 
We call $|A|^2$ the {\it transmission coefficient} and $|B|^2$ the {\it reflection coefficient}. 
Along with the passage from the left to the right of the support of $V$, consider the passage from the right to 
the left. The corresponding solution $v$ of (\ref{2}) is 
$e^{-ikx/h}+B_2 e^{ikx/h}$ to the right of $\supp V$, and $A_2 e^{-ikx/h}$ to the left. 
The {\it scattering matrix} is defined as 
$$S(k)=\begin{pmatrix}A&B\\ -\overline BA/\overline A&A\end{pmatrix}\in \U(2)$$
$S(k)$ remains unitary and symmetric for complex values of $k$.  
{\it Resonances} of (\ref{2}) are then defined as $E=k^2\in{\bf C}$, where $k$ is a pole of $S$, and {\it physical resonances}
those with $\im k>0$. Thus $E$ is a resonance iff the solution of (\ref{3}) is purely outgoing as $x\to+\infty$ and $x\to-\infty$. 
The poles coincide with the poles of meromorphic extension of the resolvent $(P-k^2)^{-1}$ from the physical
half-plane $\im E<0$ to the second sheet $\im E>0$.\\

\noindent {\it b) Monodromy matrix for BdG equation: heuristics.}

\smallskip
Now we discuss BdG equation $({\cal P}(x,hD_x)-E)U=0$ for large $|x|$, 
i.e. (within our approximation above) when $|x|\geq L+\ell$, so $\Delta(x)=\Delta_0$, $\mu(x)=\mu_0>E$. Solutions are of the form
$$U(x;h)=\begin{pmatrix}a&b\\c&d\end{pmatrix}{e^{ikx/h}\choose e^{i\ell x/h}}$$ 
$\mu_0+E\pm i\Delta_0\in\{k^2,\ell^2\}$, so eigenfrequencies are $(\pm k,\pm\overline k)$,
$k=\sqrt{\mu_0+E+i\Delta_0}$, and the corresponding solutions as follows:

Let $\phi(x)=\sgn(x)\phi$, ${\cal Z}$ be the 2-D complex line bundle spanned by $F_1^\pm(x)={e^{i\phi(x)/2}\choose -i} e^{\pm ikx/h}$ 
(associated with the scattering process
$e^+\to e^-$), and 
$\overline {\cal Z}$ the 2-D complex line bundle spanned by $F_2^\pm(x)={e^{i\phi(x)/2}\choose i} e^{\pm i\overline kx/h}$
(associated with the scattering process $e^-\to e^+$). 

The space of solutions of exponential type for BdG is
${\cal Z}\oplus\overline{\cal Z}$, and ${\cal Z},\overline{\cal Z}$ are orthogonal for the usual pointwise Hermitian product in ${\bf C}^2$. 
Declare that $E\in{\bf C}$ is a ${\cal Z}$-{\it resonance}
iff the ${\cal Z}$-component of the wave function solving BdG equation is outgoing and evanescent (``physical solution'') at infinity, i.e.
\begin{eqnarray*}
&U(x,h)=A{e^{i\phi/2}\choose -i} e^{ikx/h}, x\to+\infty\\
&U(x,h)=B{e^{-i\phi/2}\choose -i} e^{-ikx/h}, x\to-\infty
\end{eqnarray*}
Similarly we say that $E$ is a $\overline{\cal Z}$-{\it resonance} iff the
$\overline{\cal Z}$-component of the wave function is outgoing (and evanescent) at infinity, i.e.  
\begin{eqnarray*}
&U(x,h)= A{e^{i\phi/2}\choose i} e^{-i\overline kx/h}, x\to+\infty\\ 
&U(x,h)= B{e^{-i\phi/2}\choose i} e^{i\overline kx/h}, x\to-\infty
\end{eqnarray*}
So for both sets of resonances, the corresponding solution is simultaneously decaying, and outgoing at $\pm\infty$.
These sets of resonances need not coincide (although they come up in pairs), but
their real parts are given by Bohr-Sommerfeld quantization rules. 
Namely, define the monodromy operator $M^{\cal Z}(k)$ acting on ${\cal Z}$ according to the formula
$${e^{-i\phi/2}\choose -i} e^{ikx/h}+B{e^{-i\phi/2}\choose -i} e^{-ikx/h}\mapsto A{e^{i\phi/2}\choose -i} e^{ikx/h}$$
and similarly for $M^{\overline{\cal Z}}(k)$. 
It is plausible to expect that $M^{\cal Z}(k),M^{\overline{\cal Z}}(k)\in\U(1,1)$, and that the corresponding scattering 
matrices $S^{\cal Z}(k)$, $S^{\overline{\cal Z}}(k)$ have a meromorphic 
extension to the complex plane, their poles defining the resonances $E^{{\cal Z}}$ and $E^{\overline{\cal Z}}$. 
Actually, we shall construct ``relative monodromy operators'' in the ``classically allowed region''.
In particular the relative monodromy operators are in U(1,1) for some specific Lorenzian form which is constructed below.\\ 

\section{Bohr-Sommerfeld quantization rules}
\medskip
In this work, we content to determine the real parts of the resonances, extending to this setting the method of positive commutators
elaborated in \cite{Sj}, \cite{HeSj} and \cite{IfaLouRo}. Imaginary parts may be determined as in \cite{Ro}. 
We obtain Bohr-Sommerfeld quantization rules 
for the quasi-particle, alternating even and odd quantum numbers associated with the electron and the hole. In the sequel we will sketch
a proof of the following result:\\

\noindent {\bf Theorem 1}: Let $\int_{-x_{0}}^{x_{0}}\eta^{\rho}(y;h)\,dy$ be the semi-classical actions (see Proposition 8 below)
$\rho=1$ for the electron, $\rho=-1$ for the hole. Bohr-Sommerfeld quantization conditions near $E_0$ are given at first order by:
$$\oint_{\gamma_E}\eta^\rho(y;h)\,dy-h\phi+h\pi+{\cal O}(h^2)=2\pi nh;\;\;n\in {\bf Z}$$
with even (resp. odd) quantum numbers $n$ for the electron (resp. the hole). Here $\oint_{\gamma_E}$ denotes integral over the loop $\gamma_E$
obtained by gluing together $\Lambda^>_E$ and $\Lambda^<_E$, if we ignore tunneling in momentum space.\\

\section{Microlocal solutions in Fourier representation near the branching points}
\medskip
\noindent {\it a) Reduction of the system.}

\smallskip
In $h$-Fourier representation, ${\cal F}_hu(\xi)=(2\pi h)^{-1/2}\int e^{-ix\xi/h}u(x)\,dx$ the local Hamiltonian near $a=a_E=(x_E,\xi_E)$,
${\cal P}^a$ takes the form~:
\begin{equation}
{\cal P}^a(-hD_\xi, \xi)=\begin{pmatrix}\xi^2-\mu\kern 1pt &  e^{i\phi/2}(E-\alpha hD_\xi-\alpha x_E)\kern 1pt \\
e^{-i\phi/2}(E-\alpha hD_\xi-\alpha x_E)\kern 1pt & -\xi^2+\mu\kern 1pt \end{pmatrix} 
\end{equation}
By PT symmetry ${\cal P}^{a'}={\cal I}{\cal P}^a{\cal I}$ near $a'=a'_E=(-x_E,\xi_E)$.
Solving the system ${\cal P}^a(-hD_\xi, \xi)\widehat U=0$, $\widehat U={\widehat\varphi_1\choose\widehat\varphi_2}$ gives second order ODE
for
$u(\xi)=\exp[-i\int^\xi g(s)ds/h]\widehat\varphi_2(\xi)$,
\begin{equation}
P^a(-hD_{\xi},\xi,h)u(\xi)={E^2\over\alpha^2}u(\xi)
\end{equation}
$$P^a(-hD_{\xi},\xi,h)=(hD_\xi)^2+\alpha^{-2}(\xi^2-\mu)^2+h^2(\xi^2-\mu-E )^{-2}(2\xi^2+\mu+E )$$
After $E$-dependent scalings
$\beta=\sqrt\alpha(2\xi_E)^{-3/2}>0$, $E_1=(2\xi_E)^{-2}E$,
$\xi=2\xi_E\beta\omega\xi'+\xi_E, \omega=\pm1$ 
($\xi'$ is ``local momentum'') we obtain
$P^a_{\omega}(-hD_{\xi'},\xi',h)u_{\omega}(\xi')=\bigl({E_1\over\beta}\bigr)^2u_{\omega}(\xi')$, where
$$P^a_{\omega}(-hD_{\xi'},\xi';h)=(-hD_{\xi'})^2+(\xi'+\beta\omega\xi'^2)^2+h^2\beta^2f(\omega\beta\xi')$$
is an anharmonic Schr\"odinger operator. 
The lower order term $f(z)=(2z^2+2z+{3\over4}+E_1)(z^2+z-E_1 )^{-2}$
has a pole on $\Lambda_E^>$ where the linear approximation of $\Delta(x)$
breaks down. The linear approximation only holds for small $\xi'$. Consider the map

\begin{equation}\label{iota}
\iota^a:\sum_{\omega=\pm1}\Ker _h(P^a_\omega-\bigl({E_1\omega\over\beta}\bigr)^2)\to\Ker _h({\cal P}^a-E)
\end{equation}
where $\Ker _h$  denotes the microlocal kernel. The index $\omega$ is to be chosen carefully with the complex 
germ of solutions having the right decay beyond the branching points $\pm x_E$.
We shall endow the RHS of (\ref{iota}) with a Lorenzian structure
and ``diagonalize'' $\iota^a$ in some orthogonal subspaces.\\

\noindent {\it b) The normal form of Helffer-Sj\"ostrand}

\smallskip
When $E_1<{1\over4}$, we take $P^a_\omega$ microlocally to its normal form, namely:\\

\smallskip
\noindent {\bf Proposition 2} \cite{HeSj}: There exists an analytic diffeomorphism $t\mapsto F_0(t)$ defined in a neighborhood of 0, $F_0(0)=0$, 
with inverse $G_0$, and 
a real analytic phase function $\phi_\beta(\xi',\theta)$, defined in a neighborhood of (0,0),
of the form $\phi_\beta(\xi',\theta)=\xi'\theta+g_\beta(\xi',\theta)$, $g_\beta(\xi',\theta)={\cal O}(|\xi',\theta|^3)$, 
parametrizing the canonical transformation $\kappa_\beta:(\partial_\theta \phi_\beta,\theta)\mapsto(\xi', \partial_{\xi'}\phi_\beta)$,
such that $F_0\circ p_\beta\circ\kappa_\beta=p_0$. At the semi-classical level,
there is a (formally) unitary FIO operator $A$ defined microlocally
near (0,0)
\begin{equation*}
Av(\xi',h)=(2\pi h)^{-1}\int\int e^{i\varphi(\xi',\eta,\theta)/h}c(\xi',\eta,\theta,h)e^{ib(\xi',\eta,\theta,h)}
v(\eta,h)\, d\eta d\theta
\end{equation*}
and a real valued analytic symbol 
$$F(t,\beta,h)=F_0(t,\beta)+hF_1(t,\beta)+h^2F_2(t,\beta)+\cdots$$
with $F_1(t,\beta)=-{1\over2}$ such that 
\begin{equation*}
A^*F(P_\omega,\beta,h)A=P_0(\eta,hD_\eta)={1\over2}\bigl((hD_\eta)^2+\eta^2-h\bigr), \quad A^*A\equiv\Id
\end{equation*}

The function $F_0$, taking the period $T(E)$ of Hamilton vector flow for $P^a_\omega$ at energy $(E_1/\beta)^2$ to $2\pi$,
involves an elliptic integral, which requires sometimes the use of formal calculus.\\

\noindent{\it c) Weber equation and parabolic cylinder functions}

\smallskip
Weber equation 
$P_0v=\nu hv$, through change of variables $\eta=(h/2)^{1/2}\zeta$, $\widetilde v(\zeta)=v(\eta)$ scales to
\begin{equation*}
-\widetilde v''+\frac{1}{4}\zeta^2\widetilde v=\bigl(\nu+\frac{1}{2}\bigr)\widetilde v
\end{equation*}
Fundamental solutions express as {\it parabolic cylinder} functions $D_\nu$,
entire in ${\bf C}$.
The systems $\bigl(D_\nu(\pm\zeta),D_{-\nu-1}(\pm i\zeta)\bigr)$ are fundamental solutions for any choice of $\pm$.
Integral representations
give asymptotic solutions of $(P_0-\nu h)u(\eta)=0$ by stationary phase for real $\nu$, 
${E'}_1^2=2\beta^2F(\beta^{-2}E_1^2,\beta,h)=2\beta^2(\nu+1)h$. 
\begin{eqnarray*}
&D_\nu\bigl(\eps (h/2)^{-1/2}\eta\bigr)={\Gamma(\nu+1)\over-2i\pi\sqrt h}h^{E^2/4h}
\int_\infty^{(0^+)}\exp\bigl[i\Phi^{\nu}_{\eps }(s;\eta)/h\bigr]\, ds\\
&D_{-\nu-1}\bigl(i\eps (h/2)^{-1/2}\eta\bigr)=\frac{\Gamma(-\nu)}{2i\pi}h^{-E^2/4h}
\int_\infty^{(0^+)}\exp\bigl[i\Phi^{-\nu-1}_{\eps }(s;\eta)/h\bigr]\, {ds\over s}
\end{eqnarray*}
with $\eps =\pm1$, $E=\sqrt{2(\nu+1)h}$, see \cite{Wh}. This normalization is called Whittaker normalization.
Classically forbidden regions $|\eta|>E$ lie on Stokes lines, classically 
allowed region $|\eta|<E$ in between, and  
3 Stokes lines stem from each ``turning point'' $\eta=\pm E$. \\

\noindent {\it d) Microlocal solutions.}

\smallskip
We apply asymptotic stationary phase to $AD_j$, $j\in\{\nu, -\nu-1\}$. With $h'=\beta^2 h$ as a ``rescaled'' Planck constant, we get:\\

\noindent {\bf Proposition 3}: In Fourier representation, the image $K_h^a(E)=\Ker _h({\cal P}^a(-hD_\xi, \xi)-E)$ of $\iota^a$
is a 2-D vector space spanned by the spinors
$\widehat U^j_{\eps ,\omega}={\widehat\varphi_1\choose\widehat\varphi_2}^j_{\eps ,\omega}$, $(j,\eps ,\omega)\in
\{\nu, -\nu-1\}\times\{-1,1\}^2$,
of the form:
\begin{eqnarray*}
&\widehat U^\nu_{\eps ,\omega}=C_{h'}^{\nu}\sum_{\theta_\omega=\pm\widehat\theta_\omega(\xi_1)}
{e^{i\phi/2}(\xi^2-\mu-E)^{-1/2}X^\nu_{\eps ,\omega}\choose(\xi^2-\mu-E)^{1/2}}
|\widetilde a_{\eps ,\omega}^{\nu}|\exp[i(\Phi^\nu_{\eps ,\omega}+h'R_{\omega}^{\nu})/h']+{\cal O}(h')\\
&\widehat U^{-\nu-1}_{\eps ,\omega}=C_{h'}^{-\nu-1}\sum_{\theta_\omega=\pm\widehat\theta_\omega(\xi_1)}
\eps \, \sgn(\theta_\omega){e^{i\phi/2}(\xi^2-\mu-E)^{-1/2}X^{-\nu-1}_{\eps ,\omega}\choose(\xi^2-\mu-E)^{1/2}}\\
&|\widetilde a_{\eps ,\omega}^{-\nu-1}|\exp[i(\Phi^{-\nu-1}_{\eps ,\omega}+h'R_{\omega}^{-\nu-1})/h']+{\cal O}(h')
\end{eqnarray*}
Here $\widehat\theta_\omega(\xi_1)$ is a critical point (from stationary phase), 
$\Phi^j_{\eps ,\omega}+h'R_{\omega}^{j})$ the $h'$-dependent phase functions, and $X^{j}_{\eps ,\omega}$,
$|\widetilde a_{\eps ,\omega}^{j}|$ some positive amplitudes. Spinors $U^j_{\eps ,\omega}$ verify the symmetry
${}^\dagger\widehat {U}^{j}_{-\eps ,-\omega}=\widehat {U}^{j}_{\eps ,\omega}$
for the ``local time'' reversal operator ${}^\dagger u(\xi_1)=u(-\xi_1)$, and
the constants $C_{h'}^j$ (from Whittaker normalization of $D_\nu$, $D_{-\nu-1}$) are related by 
$C_{h'}^{\nu}C_{h'}^{-\nu-1}=\bigl((2\sqrt{h'})^3\pi^2\sin\pi\nu\bigr)^{-1}$. \\

\section{Normalization}
\medskip
\noindent {\it a) The microlocal Wronskian.}

\smallskip
We extend to BdG Hamiltonian the classical ``positive commutator method'' using conservation of some quantity 
called a ``quantum flux' (\cite{Sj}, \cite{HeSj}, \cite{Ro}, \cite{IfaLouRo}). \\

\noindent{\bf Definition 4}: Let ${\cal P}$ be (formally) self-adjoint, and $U^a,V^a\in K_h(E)$ be supported on 
$\Lambda_E^>$. We call the sesquilinear form ${\cal W}^a_\rho(U^a,V^a)=\bigl({i\over h}[{\cal P},\chi^a]_\rho U^a|V^a\bigr)=
\bigl({i\over h}[{\cal P},\chi^a]_\rho \widehat U^a|\widehat V^a\bigr)$
the {\it microlocal Wronskian} of $(U^a,\overline{V^a})$ in $\omega^a_\rho$. 
Here ${i\over h}[{\cal P},\chi^a]_\rho$ denotes the part of the 
commutator supported microlocally on $\omega^a_\rho$ (a small neighborhood
of $\supp [{\cal P},\chi^a]\cap\Lambda_E$ near $\rho$).  \\

A crucial property of the microlocal Wronskian is to be invariant by Fourier transformation: ${\cal W}^a_\rho(U^a,V^a)=
{\cal W}^a_\rho(\widehat U^a,\widehat V^a)$. The relation
${\cal W}^a_+(U^a,V^a)+{\cal W}^a_-(U^a,V^a)=0$ doesn't readily follow as in the scalar case \cite{IfaLouRo}, 
the microlocal solutions being neither smooth in spatial of Fourier
representation near the branching point, but from a careful inspection, involving also
formal calculus. This is used essentially in Propositions 5 and 8 below. 
Choosing $\eps ,\omega$ such that $\eps \omega=1$ we define a Lorenzian
metric ${\cal W}_\rho$ on the space of microlocal solutions near $a$.
In the basis $\widehat U^{j}_{\eps ,\omega}, j\in\{\nu,-\nu-1\}$ we have, up to a constant factor:
\begin{equation*}
\rho{\cal W}_{\rho}=
\begin{pmatrix}|C_{h'}^{\nu}|^2{\cal O}(h')&
C_{h'}^{\nu}\overline{C_{h'}^{-\nu-1}}\exp[-i\pi{E'_1}^2/4h']\bigl(1+{\cal O}(h')\bigr)\\
\overline{C_{h'}^{\nu}}C_{h'}^{-\nu-1}\exp[i\pi{E'_1}^2/4h']\bigl(1+{\cal O}(h')\bigr)&
|C_{h'}^{-\nu-1}|^2{\cal O}(h')\end{pmatrix}
\end{equation*}
Changing Whittaker normalization for the $D_\nu,D_{-\nu-1}$ functions, and the microlocal solutions by some 
constant phase factors, we can reduce to
$\rho{\cal W}_{\rho}=\begin{pmatrix}0&1\\1&0\end{pmatrix}+{\cal O}(h')$, and prove:\\

\noindent {\bf Proposition 5}: Under PT symmetry above
the microlocal Wronskians ${\cal W}^{a}_\rho$ endow $K_h^a(E)$ (mod $h'$) with a Lorenzian form ${\cal W}^a={1\over2}({\cal W}^a_+-{\cal W}^a_-)$. 
The same holds at $a'$, and the corresponding structures on $K_h^{a}\times K_h^{a*}$ and $K_h^{a'}\times K_h^{a'*}$
are anti-isomorphic.
The group of automorphisms preserving ${\cal W}^{a}$ and ${\cal W}^{a'}$ mod ${\cal O}(h')$ is therefore U(1,1).\\

\section{Spinors in the spatial representation}
\medskip
We compute  $U^{a,j}_{\eps ,\omega},U^{a',j}_{\eps ,\omega}$ in spatial representation, then extend along the branches $\rho=\pm1$
of $\Lambda_E^>$ with WKB solutions. \\

\noindent {\it a) Spinors near the branching points.}

\smallskip
Near $a,a'$ we apply inverse $h$-Fourier transform and get:\\

\noindent {\bf Proposition 6}: Up to a constant phase factor
\begin{eqnarray*}
&{U}^\nu_{\eps ,\omega}(x,h)=2\omega\beta\xi_E e^{ix\xi_E/h}
\sum_{\rho=\pm}{e^{i\phi/2}(\xi^2-\mu-E)^{-1/2}X^\nu_{\eps ,\omega}\choose(\xi^2-\mu-E)^{1/2}}
|a_{\eps ,\omega}^\nu|\big|_{\theta_1=\theta_\omega(\xi_1),\xi_1=\xi_\omega^\rho(x)}\\
&\times\bigl(\frac{L_\omega^\rho(x)}{i}\bigr)^{-1/2}
\exp[i\bigl(\Psi^{\nu,\rho}_{\eps ,\omega}(x)+h'R^{\nu,\rho}_{\eps,\omega}(x)\bigr)/h'](1+{\cal O}(h'))\\
&{U}^{-\nu-1}_{\eps ,\omega}(x,h)=2\omega\beta\xi_E e^{ix\xi_E/h}
\sum_{\rho=\pm} \eps\, \sgn(\theta_1)
{e^{i\phi/2}(\xi^2-\mu-E)^{-1/2}X^{-\nu-1}_{\eps ,\omega}\choose(\xi^2-\mu-E)^{1/2}}
|\widetilde a_{\eps ,\omega}^{-\nu-1}|\big|_{\theta_1=\theta_\omega(\xi_1),\xi_1=\xi_\omega^\rho(x)}\\
&\times\bigl(\frac{L_\omega^\rho(x)}{i}\bigr)^{-1/2}
\exp[i\bigl(\Psi^{\nu,\rho}_{\eps ,\omega}(x)+h'R^{\nu,\rho}_{\eps,\omega}(x)\bigr)/h'](1+{\cal O}(h'))\\
\end{eqnarray*}
Here $\bigl(L_\omega^\rho(x)\bigr)^{-1/2}$ is a real density (singular at $x=x_E$), 
and $\rho$ labels the branch of the Lagrangian manifold. The phases $\Psi^{j,\rho}_{\eps ,\omega}(x)+h'R^{j,\rho}_{\eps,\omega}(x)$,
$j\in\{\nu,-\nu-1\}$ differ only by a constant. \\


\noindent {\it b) WKB spinors away from the branching points}

\smallskip
The Lagrangian manifold $\Lambda_E^>$ consists of 2 branches $\Lambda_E^{>,\rho}$ (or simply $\rho$)
$\rho=\pm1$ so that $\rho=+1$ 
belongs to the electronic state ($\xi_1>0$ in the local coordinates near $a$ above), resp. $\rho=-1$ to the hole state ($\xi_1<0$).
These states mix up when $\Delta(x)\neq0$, but we can sort them out semiclassically, 
outside $a,a'$. 
Call the vector space of ${\bf C}^2$ generated by ${1\choose0}$ the space of (pure) {\it electronic states}, 
or {\it electronic spinors}, and this by ${0\choose1}$  the space of 
(pure) {\it hole states}, or {\it hole spinors}. 

The principal symbol ${\cal P}(x,\xi)$ has eigenvalues
$\lambda_\rho=\rho\lambda(x,\xi)=\rho\sqrt{\Delta(x)^2+(\xi^2-\mu(x))^2}$.
By diagonalizing, we obtain a line bundle
$\Lambda_E^\rho$ with fiber
\begin{equation*}
Y_\rho(x,\xi)=(\Delta^2+(-\xi^2+\mu+\rho\sqrt{\Delta^2+(\xi^2-\mu)^2}\,)^2)^{-1/2}
{\Delta e^{i\phi/2}\choose
-\xi^2+\mu+\rho\sqrt{\Delta^2+(\xi^2-\mu)^2}}
\end{equation*}
Looking at the electronic state, we choose $\rho=+1$ so that
$\lambda_\rho(x_\rho,\xi_\rho)-E=0$, while $\lambda_{-\rho}(x_\rho,\xi_\rho)-E$ is elliptic.
and similarly when looking at the hole state.\\

\noindent {\bf Proposition 7} The microlocal kernel $\Ker_h({\cal P}-E)$ on $\Lambda_E^{>,\rho}$ is one-dimensional space spanned by
\begin{equation*}
W^\rho(x,h)= e^{iS_\rho(x,h)/h}\bigl(w^\rho_0(x,h)Y_\rho(x,\partial_{x}S_\rho)+{\cal O}(h)\bigr)=
e^{iS_\rho(x,h)/h}\widetilde W^\rho(x,h)
\end{equation*}
where $w_0^\rho(x)|dx|^{1/2}$ is a smooth half-density. By the uniqueness property of WKB solutions along simple bicharacteristics,
the $h$ (or $h'$)-dependent phase function $S_\rho(x,h)$ should coincide, up to a constant (in a punctured neighborhood of $a$) with 
either one of $\Psi^{j,\rho}_{\eps ,\omega}(x)+h'R^{j,\rho}_{\eps,\omega}(x)$ above, $j\in\{\nu,-\nu-1\}$, and similarly for the half-densities.\\

\section{Relative monodromy matrices}

Now we look for connexion formulas.
For each $\eps ,\omega, \rho=\pm1, j\in\{\nu,-\nu-1\}$, 
the normalized microlocal solutions $U^{a',j,\rho}_{\eps ,\omega}$ are related to the extension $U^{a,k,\rho}_{-\eps ,-\omega,\ext }$
of the normalized microlocal solutions $U^{a,k,\rho}_{\eps ,\omega}$ along the bicharacteristics by a monodromy matrix 
\begin{equation*}
{\cal M}^{a,a',\rho}=\begin{pmatrix}d_{11}^\rho&d_{12}^\rho\\ d_{21}^\rho&d_{22}^\rho\end{pmatrix}\in U(1,1)
\end{equation*} 
(defined at least mod ${\cal O}(h')$) which we call a {\it relative monodromy matrix}. 
Since there is a pair of particles, the symmetry between the ${\cal M}^{a,a',\rho}$ and ${\cal M}^{a',a,\rho}$ is order 4; 
${\cal M}^{a',a,\rho}\in U(1,1)$ is obtained by extending from the left to the right, and applying symmetry 
\begin{equation}\label{5}
\rho{\cal M}^{a',a,\rho}={\cal I}({\cal M}^{a,a',\rho})^{-1}{\cal I}=, \quad \rho=\pm1
\end{equation}
where ${\cal I}$ denotes complex conjugation.
We compute the coefficients $d_{ij}=d_{ij}^\rho$.  
Considering behavior of $U^{a',j,\rho}_{\e ,\omega}$ in the classically forbidden region (according to scattering
process $e^+\to e^-$ or $e^-\to e^+$)  we obtain 
\begin{equation*}
{\cal M}^{a,a',\rho}=\begin{pmatrix}0&d_{12}^\rho\\ d_{21}^\rho &0\end{pmatrix}, \quad \overline{d_{12}^\rho}\, d_{21}^\rho=1
\end{equation*}
Note that if we do not look too closely at the relevant complex branches, as is the case when computing BS, it makes no difference to choose
instead ${\cal M}^{a,a',\rho}=\begin{pmatrix}d_{11}^\rho&0\\0&d_{22}^\rho\end{pmatrix}$, with $\overline{d_{11}}^\rho\, d_{22}^\rho=1$. 

As in \cite{Sj}, \cite{HeSj}, \cite{Ro}, \cite{IfaLouRo}, the argument consists now in extending microlocal solutions obtained above
from $a$ to $a'$, and computing the resulting semi-classical action. 
So take first $U_1$ equal to $U_1^a=U^{\nu,a}_{\eps ,\omega}$ near $a$, extend it along to $a'$ along the bicharacteristics $\rho=\pm1$
by WKB. Evaluating on $\rho$ near $a'$ we find $U_1^{a',\rho}=U^{\nu,a,\rho}_{\eps ,\omega,\ext}=d_{21}^\rho U^{-\nu-1,a',\rho}_{\eps ,\omega}$. 
Similarly, take $U_2$ starting at $a'$ and with $-\nu-1$ instead of $\nu$, we get
$U_1^{a,\rho}=U^{-\nu-1,a',\rho}_{\eps ,\omega,\ext}=e_{12}^\rho U^{\nu,a,\rho}_{\eps ,\omega}$, where $e_{12}^\rho=\rho\bigl(d_{21}^\rho\bigr)^{-1}$ 
is the matrix element of ${\cal M}^{a',a,\rho}$ given in (\ref{5}). We compute $d_{21}^\rho$ in two different ways and compare the result. 

(1) Using time-reversal and PT symmetries in the microlocal Wronskians, we get
\begin{eqnarray*}\label{6}
&\big(\frac{i}{h}\,[\mathcal{P}^{a'}, \chi^{a'}]_\rho U_{1}|U^{\nu}_{\eps ,\omega}\big)=d_{21}^\rho\,
\big(\frac{i}{h}\,[\mathcal{P}^{a'}, \chi^{a'}]_\rho U_{\eps ,\omega}^{-\nu-1}|U^{\nu}_{\eps ,\omega}\big)=\\
&=d_{21}^\rho\,\mathcal{W}^{a'}_\rho\big(U_{\eps, \omega}^{-\nu-1},U_{\eps ,\omega}^{\nu}\big)
=d_{21}^\rho\,\mathcal{W}^{a'}_\rho\big(\widehat{U}_{\eps ,\omega}^{-\nu-1},\widehat{U}_{\eps ,\omega}^{\nu}\big)=\\
&=-d_{21}^\rho\,\overline{\mathcal{W}^{a}_\rho\big(\widehat{U}_{-\eps ,-\omega}^{-\nu-1},\widehat{U}_{-\eps ,-\omega}^{\nu}\big)}=
-d_{21}^\rho\,\mathcal{W}^{a}_\rho\big(\widehat{U}_{-\eps ,-\omega}^{\nu},\widehat{U}_{-\eps ,-\omega}^{-\nu-1}\big)=-d_{21}^\rho\\
\end{eqnarray*}
(2) Using the extensions described in Proposition 7. Near $a'$ we have $U^\rho_{1,\ext}=e^{i\phi/2}W^\rho(x,h)=d_{21}^\rho U_{\eps ,\omega}^{-\nu-1,a',\rho}$ 
(by solving transport equation along $\rho$ the amplitude picks up the phase factor $e^{i\phi/2}$), so we need to compute
$\big(\frac{i}{h}\,[\mathcal{P}^{a'}, \chi^{a'}]_\rho W^\rho(x,h)|U^{\nu}_{\eps ,\omega}\big)$. The amplitude 
$W^\rho(x,h)$ is actually defined up to a real, constant factor $\widetilde{C}^{\rho}$.\\

\noindent {\bf Proposition 8}: Let 
$\widetilde{\Psi}^{\nu,a',\rho}_{\eps ,\omega}(x)=x\,\xi_{E}+\frac{(2\,\xi_{E})^{3}}{\alpha}\,\Psi^{\nu,a',\rho}_{\eps ,\omega}(x)$. We have
\begin{equation}\label{wr}
\big(\frac{i}{h}[\mathcal{P}^{a'},\chi^{a'}]_{\rho}W^{\rho}|U^{\nu,a',\rho}_{\eps ,\omega}\big)=
2\,\widetilde{C}^{\rho}\,e^{i\pi/4}\int \exp\bigl[i\big(\widetilde S_\rho(x;h)/h\bigr]\beta(x,h)\,(\chi_{1}^{a'})'(x)\,dx
\end{equation}
where the amplitude $\beta(x,h)$, real mod ${\cal O}(h)$, is computed from the WKB solutions in Proposition 7, and 
\begin{eqnarray*}
&\widetilde S_\rho(x,h)=S_\rho(x;h)-\bigl(x\xi_E+\widetilde{\Psi}^{\nu,a',\rho)}_{\eps ,\omega}(x)
-h\,R^{\nu}_{-\omega}\big(\theta_{-\omega}(\xi_{-\omega}^{\rho}(-x))\big)=\\
&\frac{(2\,\xi_{E})^{3}}{\alpha}\,\Psi^{\nu,a,\rho}_{-\eps ,-\omega}(x_{0})-
\int_{-x_{0}}^{x_{0}}\eta^{\rho}(y;h)\,dy+h\,R^{\nu}_{-\omega}\big(\theta_{-\omega}(0)\big)
\end{eqnarray*}
Moreover, $\beta(x,h)$ is also independent of $x$, 
so that, comparing the former expression (1) and (\ref{wr}) for a suitable choice of $\widetilde{C}^{\rho}$, we get
\begin{equation}
d_{21}^{\rho}=-e^{i\tau^{\rho}(h)/h}\,\int(\chi_{1}^{a'})'(x)\,dx=e^{i\tau^{\rho}(h)/h}
\end{equation}
Here $\tau^{\rho}(h)=h\,\frac{\phi}{2}+h\,\frac{\pi}{2}-\int_{-x_{0}}^{x_{0}}\eta^{\rho}(y;h)\,dy+\Const $,
where $\Const $ is evaluated at the boundaries $x=\pm x_E$, and depends only on $E'_1$. 
It will eventually disappear from the final formula, by adding to BS the contribution of the lower branch $\Lambda^{<,\rho}_E$.
Note that $\int_{-x_{0}}^{x_{0}}\eta^{\rho}(y;h)\,dy$,
$\eta^\rho(y;h)$ being the derivative of the $h'$-depending phase function, is the semi-classical action.

\section{Bohr-Sommerfeld quantization rules}

We set $F^{j,a,\rho}_{\eps ,\omega}=\frac{i}{h} [\mathcal{P}^{a},\chi^{a}]_{\rho}U^{j,a,\rho)}_{\eps ,\omega}$, and similarly with $a'$.
The set $\{G^{j,\flat}_{\eps ,\omega}=F^{j,\flat,+}_{\eps ,\omega}-F^{j,\flat,-}_{\eps ,\omega}: j\in\{\nu, -\nu-1\}, \flat\in\{a,a'\}\}$ 
(or their $h$-Fourier transform)
can be interpreted as a basis of the microlocal co-kernel of ${\cal P}$ near $a,a'$. 
Following \cite{IfaLouRo}, we introduce Gram matrix $\mathcal{G}^\rho$ of vectors
$\widehat{U}_{1}^\rho$ and $\widehat{U}_{2}^\rho$ in this basis, namely
$\mathcal{G}=\begin{pmatrix}
\big(\widehat{U}_{1}|\widehat G^{-\nu-1,a}_{\eps ,\omega}\big)&
\big(\widehat{U}_{2}|\widehat G^{-\nu-1,a}_{\eps ,\omega}\big) \\
\big(\widehat{U}_{1}|\widehat G^{\nu,a'}_{\eps ,\omega}\big) & 
\big(\hat{U}_{2}|\widehat G^{\nu,a'}_{\eps ,\omega}\big) \end{pmatrix}$. Using symmetries  we get
$$\mathcal{G}=\mathcal{G}^\rho=2\begin{pmatrix}1&e^\rho_{12}\\-d^\rho_{21}&-1\end{pmatrix}$$
The condition $\det(\mathcal{G}^{(\rho)})=0$ means that $U_1$ is colinear to $U_2$, i.e. there is a global section 
of $\Ker_h(\mathcal{P}-E)$. Recall $e_{12}^\rho=\rho\bigl(\overline{d_{21}^\rho}\bigr)^{-1}$; for $\rho=+1$ (electronic state)  
we get $\im d_{21}^+=0$, that is $\sin\big(\frac{\tau^{(+)}(h)}{h}\big)=0$. We eventually obtain BS by ``surgery'':
namely (ignoring tunneling) we cut and paste the half-bicharacteristic $\Lambda^{>,+}_E$ 
in the upper-half plane $\xi>0$ with its symmetric part $\Lambda^{<,-}_E$ in $\xi<0$
and add together the contributions. By symmetry, the constant term $\Const $ in $\tau^+(h)$ drops out, while the other terms
$h\,\frac{\phi}{2}+h\,\frac{\pi}{2}-\int_{-x_{0}}^{x_{0}}\eta^{\rho}(y;h)\,dy$ add up, which yields BS for the electronic state. 
We argue similarly for the hole state. This eventually gives Theorem 1.\\

\noindent {\it Acknowledgements}: We thank Timur Tudorovskiy for having introduced us to the problem. This work has been partially supported by
the grant PRC CNRS/RFBR 2017-2019 No.1556 ``Multi-dimensional semi-classical problems of Condensed Matter Physics and Quantum Dynamics''.\\


\begin{thebibliography}{9}


\bibitem{Ar2} Arnold V 1983 {\it Geometrical methods in the theory of ordinary differential equations} (Springer, Berlin)

\bibitem{BCS} Bardeen J, Cooper L and Schriefer J 1959 {\it Phys. Rev.} {\bf 108(5)} 1175

\bibitem{BenMhRo} Bensouissi A, M'hadbi N and Rouleux M 2011 
{\it Proc. ``Days of Diffraction 2011'' (Saint-Petersburg)} (IEEE 101109/DD.2011.6094362) 39

\bibitem{CaMo} Cayssol J and Montambaux G 2004 
{\it Phys.Rev.B} {\bf 70} 224520

\bibitem{ChLeBl} Chtchelkatchev N, Lesovik G and Blatter G 2000 {\it Phys.Rev.B} {\bf 62(5)} 3559

\bibitem{deGe} de Gennes P G 1966 {\it Superconductivity of Metals and Alloys} (Benjamin, New York)

\bibitem{GeSi} G\'erard C and  Sigal I M 1992 {\it Comm. Math. Phys.} {\bf 145} 281 

\bibitem{DuGy} Duncan K P and Gy\"orffy B L 2002 {\it Annals of. Phys.} {\bf 298} 273

\bibitem{HeSj} Helffer B and Sj\"ostrand J 1989 {\it Soc. Math. de France, M\'emoire 39} {\bf 117(4)}

\bibitem{IfaLouRo} Ifa A, Louati H and Rouleux M 2018 {\it J. Math. Sci. Univ. Tokyo} {\bf 25} 1

\bibitem{Ro} Rouleux M 1999 {\it Tunneling effects for h-Pseudodifferential Operators, Feshbach resonances and the Born-Oppenheimer 
approximation } (Adv. Part. Diff. Eq. {\bf 16}) ed M Demuth, E Schrohe {\it et al.}
(Wiley VCH, Berlin) 

\bibitem {Sj} Sj\"ostrand J 1990 Density of states oscillations for magnetic Schr\"odinger operators  
{\it Proc. Diff. Eq. Math. Phys. (Univ. Alabama, Birmingham)} ed Bennewitz 

\bibitem{Wh} Whittaker E T and Watson G N 1980 {\it A Course of Modern Analysis} (Cambridge Univ.Press)

\end{thebibliography}
\end{document}